# High Resolution Isotropic 3D Cine imaging with Automated Segmentation using Concatenated 2D Real-time Imaging and Deep Learning

Mark Wrobel[a], Michele Pascale[a], Tina Yao[a], Ruaraidh Campbell[a], Elena Milano[b], Michael Quail[a,b], Jennifer Steeden[a], Vivek Muthurangu[a]

[a] UCL Centre for Translational Cardiovascular Imaging, University College London, 20c Guilford St, London, UK, WC1N 1DZ

[b] Great Ormond Street Hospital, Great Ormond St, London, UK, WC1N 3JH

Joint senior authors: V.M. and J.S.

Address correspondence to: V.M. (email: v.muthurangu@ucl.ac.uk)

# Abstract

**Background**: Conventional cardiovascular magnetic resonance (CMR) in paediatric and congenital heart disease uses 2D, breath-hold, balanced steady state free precession (bSSFP) cine imaging for assessment of function and cardiac-gated, respiratory-navigated, static 3D bSSFP whole-heart imaging for anatomical assessment. Our aim is to concatenate a stack 2D free-breathing real-time cines and use Deep Learning (DL) to create an isotropic a fully segmented 3D cine dataset from these images.

**Methods:** Four DL models were trained on open-source data that performed: a) Interslice contrast correction (de-banding); b) Interslice respiratory motion correction; c) Super-resolution in the slice direction; and d) Segmentation of right and left atria and ventricles (RA, LA, RV, and LV), thoracic aorta (Ao) and pulmonary arteries (PA). In 10 patients undergoing routine cardiovascular examination, our method was validated on prospectively acquired sagittal stacks of real-time cine images that were converted to segmented, isotropic 3D cine data. Quantitative metrics (ventricular volumes and vessel diameters) and image quality of the 3D cines were compared to conventional breath hold cine and whole heart imaging.

**Results:** All real-time data were successfully transformed into 3D cines with a total offline reconstruction and post-processing time of <1 min in all cases. There were no significant biases in any LV or RV metrics (bias ± standard deviation in ml, LV EDV: 0.4 ± 7.7, LV ESV: 1.3 ± 6.5, RV EDV: -4.4 ± 9.6, RV ESV: 3.7 ± 14.5) with reasonable limits of agreement and correlation. There is also reasonable agreement for all vessel diameters, although there was a small but significant overestimation of RPA diameter (0.8 mm, p-value = 0.0037). The DL processed 3D cine data was assessed to be of diagnostic quality unlike the unprocessed data.

**Conclusion:** We have demonstrated the potential of creating a 3D-cine data from concatenated 2D real-time cine images using a series of DL models. Our method has short acquisition and reconstruction times with fully segmented data being available within 2 minutes. Our models are trained from fully open-source datasets, allowing our technique to be easily shared with other clinical centres. The good agreement with conventional imaging suggests that our method could help to significantly speed up CMR in clinical practice.

# Introduction

Cardiovascular magnetic resonance (CMR) is heavily used in the evaluation of paediatric and congenital heart disease (PCHD), providing reference-standard assessment of: i) Cardiac volumes and function using two-dimensional (2D), breath-hold, balanced steady state free precession (bSSFP) cine imaging [1,2]; and ii) Heart and vessel anatomy using cardiac-gated, respiratory-navigated, static 3D bSSFP whole-heart (WH) imaging.

More recently, 3D bSSFP WH cine techniques (3D-cine) have been developed, providing both functional and anatomical information in one easy-to-plan scan [3, 4] . However, 3D-cine imaging has poor myocardial-blood pool contrast when compared to conventional 2D cine imaging [3]. Contrast agents (particularly blood pool agents) can be used to improve image quality, but they add expense and complexity to protocols. An alternative approach is to simply concatenate a stack of breath-hold 2D cines in the slice direction, producing a 'psuedo' 3D-cine dataset with high myocardial-blood pool contrast. Unfortunately, there are several problems with this approach including: i) The need for multiple breath-holds, which can be difficult in children and take significant time to acquire; ii) Low-resolution in the slice direction, preventing proper evaluation of dynamic 3D anatomy; iii) Slice misalignment due to inconsistent breath-hold position; and iv) Image banding due each slice being acquired independently.

A solution to the multiple breath-holds problem is real-time imaging, which has been shown to be a robust alternative to breath-hold cine approaches [5]. Real-time imaging enables a stack of 2D cines to be acquired during free-breathing in less than a minute. In addition, the use of Deep Learning (DL) reconstruction for real-time 2D cines provides near breath-hold image quality with very short reconstruction times [5,6]. In this study, we propose to concatenate a stack of rapidly acquired, DL reconstructed, 2D real-time bSSFP cines to produce a pseudo 3D-cine dataset. We then use DL to sequentially remove image banding, correct slice misalignment, and super-resolve in the slice direction. This overcomes the other problems of a concatenated multi-slice approach and enables rapid acquisition and reconstruction of isotropic 3D-cine data.

We also leveraged DL to overcome another problem with 3D-cine imaging, namely time-consuming and complex post-processing. Specifically, DL is used to provide 3D+time segmentation of the cardiac chambers and great vessels for automatic quantification of atrial size, ventricular volumes and aortic and pulmonary artery diameters.

The general aim of this work was to develop a free-breathing 3D-cine technique that takes less than 1 minute to acquire and less than 1 minute to process and segment. The specific aims of this study are: i) Develop DL models for both creation and segmentation of 3D-cine using open-source data for training; ii) Compare ventricular volumes and function derived from our automatically segmented 3D-cine data to conventional manually segmented short axis breath-hold 2D cines; and iii) Validate great vessels diameters from our automatically segmented 3D-cine data to conventional manually segmented WH imaging.

## Methodology

### Overview

In this pipeline, we acquired a sagittal stack of 2D free-breathing real-time cine images (reconstructed using DL as previously described [6]), which were transformed into a fully segmented isotropic 3D whole-heart cine dataset. This transformation was achieved by sequentially applying four DL models that performed: a) Interslice contrast correction (de-banding); b) Interslice respiratory motion correction; c) Super-resolution in the slice direction; and d) Segmentation of right and left atria and ventricles (RA, LA, RV, and LV), thoracic aorta (Ao) and pulmonary arteries (PA). All DL models were trained using TensorFlow (version 2.10.1; http://www.tensorflow.org), Keras (version 2.10.0) and TensorFlow MRI (version 0.23.0) [7,8]. All code is available on GitHub (https://github.com/mrphys/3D-Cine).

### Training Data

*Training data for de-banding, respiratory-correction and super-resolution models*

The training data consisted of 111 cardiac-gated, respiratory-navigated, 3D whole-heart scans from two publicly available datasets: i) Multi-Modality Whole-Heart Segmentation (MMWHS, n = 60); and ii) Whole-Heart and Great Vessel Segmentation from 3D Cardiovascular MRI in Congenital Heart Disease (HVSMR, n = 51). It should be noted that the original HVSMR dataset contains 60 scans but 9 were excluded due to very small hearts. All volumes were rescaled to 1.5 mm isotropic resolution using bi-cubic interpolation and cropped or padded to a matrix of 256x128x80 (Head-foot, Anterior-Posterior, and Left-Right directions). All underlying volumes were processed with Contrast Limited Adaptive Histogram Equalization (CLAHE) to improve consistency of blood pool contrast. These high-resolution (HighRes) volumes were augmented 10x by rotations (randomly up to ±10 degrees in each direction) to produce 1110 datasets for training. Augmentation was performed prior to training rather than on-the-fly because simulation of respiratory motion (described below) during training would have been too time-consuming.

These volumes were each processed to mimic a single time frame of the prospectively acquired stack of sagittal real-time 2D cines, which involved simulating: i) Lower resolution in the slice (Left-Right) direction; ii) Independent respiratory motion in each slice; and iii) Interslice contrast differences. Thick slices were simulated by averaging across four adjacent slices of the HighRes data to create a low-resolution (LowRes) volume, with a simulated slice thickness of 6.0 mm. Simulation of respiratory motion was more complex and first required creation of a normalized breathing signal for each volume in the training dataset, as shown in equation 1:

$$S(t) = M \cdot Var_1(resp\ cycle) \cdot \sin\left(\frac{2\pi t}{B * Var_2(resp\ cycle)} + \varphi\right) \quad (1)$$

where *S(t)* was the respiratory signal over the total simulated acquisition time *(t)* as shown in Supplementary Figure 1. The signal was based on a sine wave, which was randomly parametrized for each volume as follows: *M* was the magnitude of breathing (range 0.5-1.5), *Var$_1$(resp cycle)* was a parameter (between 0.9-1.1) that slightly scaled the breathing amplitude between cycles to simulate inconsistent breathing depth, *B* was the breathing period (range 3-6 s), *Var$_2$ (resp cycle)* was a parameter (between 0.95-1.05) that scaled the breathing period between respiratory cycles to simulate inconsistent breathing frequency, and *φ* was a randomly initialized phase.

To simulate the multi-slice, real-time, free-breathing 2D cine acquisition (fully described in the prospective study section) we sampled *S(t)* over a series of simulated heartbeats with the average R-R interval randomly set for each volume (range 0.6-1.2 s). In addition, random scaling was added to simulate R-R interval variation within the acquisition (range 0.95-1.05). The sampled respiratory signal amplitude during the acquisition of a given slice – *S($t_{slice}$)* - was then used to non-rigidly deform the simulated slice, as described below:

$$dy(i,j) = S(t_{slice}) \cdot f_y(i,j) \qquad (2)$$

$$dx(i,j) = S(t_{slice}) \cdot f_x(i,j)$$

where *dy(i,j)* and *dx(i,j)* were the respective y (Head-Foot) and x (Anterior-Posterior) deformations for pixel (*i,j*), which were the product of a scalar value *S($t_{slice}$)*, and the pixel dependant functions *$f_y$(i,j)* and *$f_x$(i,j)*. These pixel dependant functions were spatial parabolic curves, which create realistic deformations, centered on the middle of the image with reduced deformations toward the edge of the image. The maximum deformation of the image was 20 and 8 pixels, in the Head-Foot and Anterior-Posterior directions, respectively. Each slice in the LowRes volume was deformed with these synthetic respiratory deformations to produce the low-resolution respiratory-artefacted volume (LowRes$_{resp}$).

Finally, inter-slice contrast differences were simulated by randomly selecting slices in the LowRes$_{resp}$ volume (probability=0.5) and multiplying their intensity by a random parameter in the range 0.6-1.4. The 3D volume was then re-normalized between 0-1, creating a low-resolution volume with respiratory artefacts and contrast differences (LowRes$_{resp+band}$).

*Training data for segmentation model*

For training of the segmentation model, we used the manually segmented cardiac chamber and great vessel masks from 104 of the datasets used to train the image enhancement models described above (7 datasets were removed due to single ventricular anatomy). It should be noted that open-source segmentations were only available for 33% of the MMWHS data, and therefore additional segmentations of the cardiac chambers and great vessels were performed using 3D slicer [9] by a CMR specialist (VM >20 years' experience). The HVSMR datasets required further automated morphological image processing to include ventricular trabeculations in the blood pool to match MMWHS segmentations. Images and segmentations were rescaled to 1.5 mm isotropic resolution, cropped and processed with CLAHE (as above) and the segmentation masks were re-binarized using a threshold of <0.2. Images and segmentations were then cropped or padded to a matrix of 256x128x112 slices (Head-Foot, Anterior-Posterior, and Left-Right directions). We used 112 slices for the segmentation model instead of 80 slices (used for the image enhancement models) to ensure the whole heart was included in the volume. The image data and segmentations were augmented during training using the volumentations library (version 1.0.4) with random rotations up to ± 10 degrees in all directions and elastic transforms (see Supplementary Information 1 for full description).

**DL Architecture and Training**

The inputs and outputs to all the DL models were 3D and thus the volume at each time frame was processed independently.

*De-banding network*

The de-banding network was trained to correct inter-slice contrast differences, taking a LowRes$_{resp+band}$ volume as an input and producing a LowRes$_{resp}$ volume as an output. The network utilised a conventional 3D UNet architecture (Supplementary Figure 2) and was trained using a regularised loss. This loss combined mean absolute error (MAE) and gradient mean absolute error (GMAE) with the balance determined using Hyperband optimization (see Supplementary Information 2 for full description).

*Respiratory-correction network*

The respiratory-correction network was trained to compensate for slice misalignment due to breathing motion. The input to the model was a LowRes$_{resp}$ volume and the target was a LowRes volume (without any simulated respiratory motion). The network was based on a 3D UNet architecture (Supplementary Figure 3) with the following specific modifications: i) Slice dimension was preserved in the encoding and decoding arms through the use of anisotropic MaxPooling (4,4,1) and upsampling; ii) The outputs of the trainable portion of the network were separate *x* and *y* in-plane deformation maps for each slice in the inputted volume; and iii) Inclusion of a final non-trainable lambda layer that applied the deformation maps to the inputted volume to produce the final respiratory-corrected output. The model was trained using a combined MAE and GMAE image-based loss function (balance determined using Hyperband optimization, see Supplementary Information 2), alongside a regularization loss that minimized the L2 norm of the deformation field gradients through the in-slice plane [10] (see Supplementary Information 3 for full description).

*Super-resolution network*

The super-resolution network was trained to convert a volume with low-resolution in the slice direction to a high-resolution isotropic volume. The network was based on an asymmetric 3D UNet architecture (Supplementary Figure 4) with the following modifications: i) Preservation of the slice dimension in the shallowest two levels of the encoding arm using anisotropic MaxPooling (2,2,1), resulting in isotropic data being injected into the deepest level; ii) Isotropic MaxPooling in the deepest level before the bottle neck; and iii) Isotropic upsampling in the decoding arm at all levels, which required additional upsample layers in the skip-connections of the top levels.

Our initial training scheme used a LowRes volume as input and a HighRes volume as target. However, preliminary experimentation demonstrated sub-optimal performance with prospective data due to residual slice misalignment after respiratory-correction. Therefore, we decided to perform end-to-end training of a combined model that contained both the respiratory-correction and super-resolution networks. The benefit of this approach is that the super-resolution network also learnt to correct any residual misalignments in the respiratory-corrected input data. However, although end-to-end training produced a more robust super-resolution model, the respiratory-correction model was inferior compared to the independently trained network. Thus, at inference the independently trained respiratory-correction model was used to correct for slice misalignment, and this data was then inputted into the super-resolution network trained in an end-to-end fashion.

The end-to-end model was trained using a LowRes$_{resp}$ volume as the input and a HighRes volume as the target. The training loss was based on both the final HighRes volume output and the intermediate respiratory-corrected LowRes volume (equal weighting), with the loss at both stages incorporating MAE and GMAE (see Supplementary Information 2 for full description).

*Segmentation training data and network*

We trained a multi-class model to predict RA, LA, RV, LV, PA and Ao masks for each 3D volume. The segmentation model was based on a modified 3D UNet3+ architecture (Supplementary Figure 5). Specific modifications included: i) Full-scale encoder skip connections; ii) Encoder block depth of two and decoder block depth of one; and iii) No classification guided module (CGM) or deep supervision. The network was trained using a Focal Tversky loss alongside a surface area loss. The surface area loss counts the number of voxels on the 3D surface of the predicted and ground truth masks, and calculates the square difference of these values [11]. The inclusion of this loss improved performance compared to just the Focal Tversky loss (see Supplementary Information 4 for full description). The resulting 3D segmentations were further processed to remove or combine islands (if they formed a connected structure) from each mask.

*Training information*

Data was split 92%/8% for training and validation for the de-banding, respiratory-correction, super-resolution and segmentation models. At inference, these models were run consecutively, with each time point treated independently. The final output being an isotropic 3D-cine with corresponding four-chamber and great vessel segmentations. Training and inference of the DL models was performed on an Nvidia RTX A6000 GPU.

**Prospective study population**

Prospective data were acquired in 10 children and adults with paediatric or CHD referred to our center for clinical CMR (mean age: 24.3 ± 3.5 yr, range: 8-41 yr, male: 3) (Cardiomyopathy screen – 2, Aortopathy – 3, Tetralogy of Fallot – 3, Ebsteins anomaly – 1, Coarctation - 1). Exclusion criteria were: i) Inability to breath-hold; ii) A single ventricle; and iii) Arrhythmia. All patients were imaged on a 1.5 T MR scanner (Avanto, Siemens Medical Solutions, Erlangen, Germany) with vectorcardiogram (VCG) gating. Acquisition of additional scans was approved by the local research ethics committee and written consent was obtained from all subjects/guardians (Ref: 06/Q0508/124).

**3D-cine acquisition and post-processing**

*Creation of segmented 3D-cine data*

A contiguous sagittal stack of real-time 2D cines were acquired during free-breathing with an accelerated sorted golden angle 2D radial bSSFP single-shot sequence (TR/TE = 2.98/1.49 ms, spatial resolution ~1.5x1.5 mm, slice thickness = 6 mm, R ~ 23x, temporal resolution ~ 40 ms – interpolated to 32 frames/R-R interval, 20 to 28 slices to cover the whole heart) [6]. Deep artefact suppression-based reconstruction was performed using a previously described multi-coil complex image-based 3D UNet [6]. This reconstruction was performed online using an external GPU-enabled computer, with images sent back to the scanner using the Gadgetron framework. Acquisition of each slice required two R-R intervals, with the first R-R interval being used to reach the steady state and the second for data acquisition.

*Evaluation of Sequential Improvement in Image quality*

Subjective improvement in image quality of the 3D-cine data after application of each model was evaluated on 2D+time short axis (SAX) and four-chamber (4CH) multi-planar reformats (MPR's), as well as all the axial slices for a single diastolic frame. For each view, movie clips for each stage (uncorrected, de-banded, respiratory-corrected, and super-resolved) were displayed simultaneously in a random order to two clinical CMR specialists (MQ and EM with 14- and 8-years' experience, respectively). The clips were subjectively ranked in terms of overall image quality from best to worst (1 = best, 2 = second best, 3 = third best, 4 = worst, with the possibility of tied scoring).

*Extraction of ventricular volumes and great vessel diameters*

From the DL segmented data, ventricular volumes and great vessel measurements were automatically extracted. Volumes time curves were evaluated using the RV and LV blood pool segmentation masks across all time frames. The minimum volume was taken as the end-systolic volume (ESV) and the maximum volume as the end-diastolic volume (EDV) with ejection fraction (EF) calculated as (EDV-ESV)/EDV.

Vessel diameters were semi automatically extracted from the segmentations of the Ao and PA for the 30$^{th}$ (temporally interpolated) frame of the 3D-cine (equivalent to end-diastolic data acquisition of conventional WH scan). The masks were first meshed and skeletonized then fitted with a polynomial to create a smoother centreline (7$^{th}$ order Chebyshev was empirically chosen). The points on these centrelines corresponding to the aortic and main pulmonary artery sino-tubular junctions, as well as the midpoint of the left and right pulmonary arteries, were then manually identified for diameter measurement. From these points 64 evenly spaced radial lines were emitted orthogonal to the centrelines to measure the distance to the vessel boundary. The average vessel radius was doubled to give the final average vessel diameter.

**Conventional volumetric and anatomical assessment**

*Reference-standard multi-slice SAX 2D cine imaging*

Reference-standard assessment of ventricular volumes was performed using conventional retrospectively cardiac-gated, breath-hold, bSSFP 2D cine imaging in the short axis (TR/TE ~ 2.2/1.1 ms, spatial resolution ~ 1.3×1.3 mm, slice thickness ~ 10 mm, temporal resolution ~ 38 ms, reconstructed cardiac phases = 40). Ten to fifteen contiguous slices were acquired in the short axis to ensure coverage of the whole ventricular volume, with one slice acquired per breath-hold (~6 s). The same sequence and parameters were also used to acquire a reference-standard single slice 4CH cine.

The end-diastolic and end-systolic phases for each ventricle were chosen through visual inspection of the mid-ventricular cine and slices were manually segmented by a CMR specialist (EM). The endocardial border was traced using the OsiriX open source DICOM viewing platform

(Osirix v.9.0, OsiriX foundation, Switzerland) with the papillary muscles and trabeculae included in the blood pool mask to provide reference-standard EDV, ESV, and EF.

*Reference-standard three-dimensional whole-heart imaging*

Reference-standard anatomical evaluation was performed using a conventional cardiac-triggered, respiratory-navigated 3D whole-heart bSSFP sequence (TR/TE ~ 3.6/1.5 ms, spatial resolution ~ 1.5×1.5X1.5 mm, phase encoding direction – Anterior-Posterior, slice encoding direction – Left-Right, triggered in diastole). The median acquisition time was 227 s (range 148-845 s). Vessel diameters were measured manually by a CMR specialist (VM) on the static 3D whole-heart data using the multi-planar reformat (MPR) tool within the OsiriX viewing platform (Osirix v.9.0, OsiriX foundation, Switzerland). The vessel diameters were measured from MPR's of the aortic and main pulmonary artery sino-tubular junctions, as well as the midpoint of the left and right pulmonary arteries to provide comparison measurement for the automatically measured diameters from the 3D-cine data.

*Image quality comparison of 3D-cine with conventional 2D cine and WH imaging*

Qualitative evaluation involved scoring the image quality of the uncorrected concatenated real-time data, the fully corrected 3D-cine data, conventional 2D cine data, and 3D WH data. Specifically, 2D+time SAX and 4CH cines (from reference-standard 2D cine imaging and MPR's created from 3D-cine) and 3D WH axial stacks (from reference-standard 3D WH and derived from diastolic frame of 3D-cine) were assessed by two blinded clinical CMR specialists (MQ and EM). All clips were scored individually, in a random order on a 5-point Likert scale (5 = non-diagnostic, 4 = poor, 3 = adequate, 2 = good, 1 = excellent).

Quantitative image quality was assessed by measuring edge sharpness (ES) and myocardial-blood pool contrast. Intensity profiles were automatically extracted using segmentations of the cardiac structures to isolate the LV – septum boundary. These intensity profiles were used to calculate myocardial-blood pool contrast, by taking the ratio of the maximum (blood pool) and minimum (myocardium) intensity over the non-normalised intensity profiles. ES was calculated as the maximum gradient along the normalized intensity profile, as described in previous work [12]. ES and contrast was estimated on 4CH and SAX MPR's of the uncorrected concatenated real-time data, the fully corrected 3D-cine data, as well as for reference-standard 2D cines and MPR's 4CH and SAX MPR's of the reference-standard 3D WH data.

*Proof of concept conventional real-time sequence*

We additionally acquired a generic (non-DL) sagittal cartesian real-time stack (TR/TE ~ 2.4/1.2 ms, spatial resolution ~ 1.5×1.5 mm, slice thickness ~ 6 mm, R = TPAT3, temporal resolution ~ 110 ms, reconstructed cardiac phases = 10 and 28 slices) on a healthy volunteer. We subsequently applied our four DL models (debanding, motion correction, super-resolution, and segmentation) and visualised 4CH and SAX MPRs in order to showcase the generalisability of our method to more conventional real-time (albeit low temporal resolution) scanning.

**Statistics**

All continuous variables are presented as mean (± standard deviation) for normally distributed data, or median (interquartile range) otherwise. For distributions of multiple groups that were non-normal (Shapiro-Wilk test), a Friedman Chi Square test was used alongside a post-hoc

Nemenyi-Friedman test in order to assess any statistical differences in the metrics ($p<0.05$). For normal distributions of multiple groups, a repeated measures ANOVA was used in conjunction with a paired t-test. For paired non-normal distributions, a Wilcoxon signed-rank test was used instead.

## Results

### Model Training

The de-banding, respiratory-correction, and end to end super-resolution models took 6h, 2h, and 25h to train respectively, with improvements in SSIM, PSNR and MSE shown in Supplementary Figure 6. The segmentation model took 49h to train with validation Dice scores for the LV, RV, LA, RA, Ao, PA shown in Supplementary Table 1.

### Prospective study

Multi-slice 2D real-time cine stacks were acquired during free-breathing with an acquisition time of 42 ± 11 s (range: 26 to 65 s). This is in comparison to 341 ± 51 s (range: 275 to 444 s) for acquisition of the reference-standard breath-hold 2D cine stacks, and 311 ± 200 s (range: 158 to 845 s) for the reference-standard static 3D WH scans. All real-time data were successfully transformed into 3D-cines with the total time for de-banding, respiratory-correction and super-resolution being 7-9 s, depending on the number of slices. The segmentation model took an additional 20-25 s depending on the number of slices, resulting in a total offline reconstruction and post-processing time of <1 min in all cases.

*Sequential Improvement in Image quality*

Representative images and movies at each stage of DL reconstruction are shown in Figure 1 and Supplementary Movie 1 with image quality improving as each model is sequentially applied. This is corroborated by the qualitative image ranking, which demonstrated that as each model was applied, image ranking significantly improved (Supplementary Table 2).

*Cardiac segmentation and volume quantification*

Examples of segmentation of the cardiac chambers in both the 4CH and SAX views in systole and diastole are shown in Figure 2 (all time frames - Supplementary Movie 2). In addition, systolic and diastolic surface renders of the cardiac chambers are shown in Figure 3 (all time frames - Supplementary Movie 3), and corresponding volume curves shown in Figure 4. Visually the DL model provides robust segmentation, and this is corroborated by comparison with reference-standard volumetric analysis. Left and right ventricular volumes and ejection fraction measured using our 3D-cine and reference-standard breath-hold 2D cine sequence are shown in Table 1. There were no significant biases (0.3-6.7%) in any LV or RV metrics with reasonable limits of agreement and correlation. Corresponding Bland-Altman and scatter plots of right and left ventricular volumes (combining EDV and ESV data) are shown in Figure 5 and LVEF and RVEF in Figure 6.

*Vessel measurements quantification*

Examples of segmentation of the aorta and pulmonary arteries from the 3D-cine at systole and end diastole are shown in Figure 7 (all time frames - Supplementary Movie 4), with both systolic and diastolic surface renders of the great vessels in Figure 3 (all time frames - Supplementary Movie 5). Vessel diameters automatically measured from our 3D-cines and reference-standard static 3D WH data are shown in Table 2. There is reasonable agreement for all vessel diameters, although there was a small but significant overestimation of RPA diameter (0.8 mm, p-value = 0.0037). Figure 8 shows corresponding Bland-Altman and scatter plots for all great vessels combined.

*Image quality comparison with conventional cine and WH imaging*

The uncorrected concatenated real-time data was scored to be non-diagnostic or poor for all patients (axial stack = 5.0±0.0, 4CH and SAX MPRs = 4.6±0.5, presented as average score and standard deviation). Images derived from the 3D-cine data were generally scored as good or adequate diagnostic quality (axial stack = 2.6±0.9, MPRs = 2.7±0.7). However, reference-standard imaging scored significantly higher, consistently achieving good and excellent scores (axial stack = 1.6±0.6, MPRs = 1.0±0.0).

The results for quantitative ES are shown in Figure 9 and Table 3, showing that the final 3D-cine data had significantly higher ES than the uncorrected concatenated real-time data for both the 4CH and SAX views. There was no significant difference in ES between 3D-cine data and reference-standard 3D WH data, but the reference-standard breath-hold 2D cines did have significantly higher ES than the 3D-cine data.

The contrast results are shown in Figure 9 and Table 3. For both views, the contrast for the uncorrected concatenated real-time data and the final 3D-cine data showed no differences. For the 4CH view, the reference-standard 3D WH contrast was not significantly different to either the final 3D-cine or uncorrected concatenated real-time data however, for the SAX the WH contrast was significantly lower than both the uncorrected concatenated real-time and final 3D-cine data. For both views, the reference-standard breath-hold 2D cines showed significantly better contrast compared to all other scans.

*Proof of concept conventional real-time sequence*

Sequential DL reconstruction was performed on the conventional low temporal resolution real-time data with images and movies at each stage of DL reconstruction are shown in Figure 10 and Supplementary Movie 6. Image quality improves as each model was sequentially applied. Examples of segmentation of the cardiac chambers in both the 4CH and SAX views in systole and diastole are shown in Figure 11 (all time frames - Supplementary Movie 7). The segmentation is of good quality highlighting the generalisability of our method.

# Discussion

The main findings of this study were: i) It is feasible to use DL to de-band, respiratory-correct and super-resolve a stack of real-time 2D cines to produce an isotropic 3D-cine dataset; ii) Rapid 3D segmentation of all cardiac chambers across the cardiac cycle is possible using DL; iii) There was good agreement between our 3D-cine and conventional 2D cines for assessment of ventricular volumes and function; and iv) There was also good agreement between our 3D-cine and conventional 3D WH measurement of great vessel diameters. Importantly, our method takes less than 1 minute to acquire and less than 1 minutes to process. This makes our approach faster than conventional cine and WH imaging (>10 minutes acquisition time combined) and most current 3D-cine techniques. In addition, we have demonstrated that our method works on conventional ((but low temporal resolution) real-time data and we have made our code/models openly available. Thus, our framework can easily be used at other sites without access to cutting edge real-time imaging.

**3D-cine – Current work**

Whole-heart 3D-cine imaging has the potential to shift the paradigm in CHD evaluation, providing functional and anatomical evaluation in a single scan. Nevertheless, there are significant challenges that must be overcome to make 3D-cine imaging a clinical reality, with long acquisition time being the most pressing. Early methods such as 4D Multiphasic Steady-State Imaging with Contrast Enhancement (MUSIC) compromised on temporal resolution to ensure that scan times were <15 minutes [3]. Other 3D-cine techniques relied on both lower spatial and temporal resolution to limit scan times [13,14]. More recent 3D-cine techniques leverage significant undersampling, iterative reconstruction schemes (e.g. compressed sensing), and non-Cartesian trajectories to reduce scan time. Unfortunately, reconstruction times can range between 2.5-10 hours, which is infeasible for application in a clinical environment [15–17]. Reconstruction times can be reduced through optimization of iterative algorithms and the use of higher performance computing, but a better approach may be to use DL reconstruction. For instance, CINENet uses a highly undersampled 3D-cine acquisition alongside a 4D CNN for reconstruction and boasts a 10 second breath-hold acquisition and 5 second reconstruction [4]. This demonstrates the potential of DL to facilitate 3D-cine imaging but does not address the problem of low blood pool-myocardial contrast in 3D-cine images. To solve this problem, many studies use either Gadolinium or Ferumoxytol, but this adds cost and complexity to protocols with Ferumoxytol not being approved for use in many countries [3].

We took an alternative approach, acquiring a stack of real-time 2D SSFP cines that inherently have high contrast and can be acquired in ~40 s. This addresses the acquisition time and myocardial-blood pool contrast problems associated with most 3D-cine techniques. We then use DL to rapidly convert this data to a 3D-cine isotropic dataset (<1 minute), solving the reconstruction time problem.

**Slice-based respiratory motion correction**

Although our approach requires no contrast and can be acquired during free-breathing, it does require correction of slice misalignment to create clinically usable 3D-cine data. Conventionally, slice misalignment due to inconsistent breath-holding is corrected by registering all the slices in a stack to either a long axis image or a 3D volume (slice-to-volume registration - SVR). These techniques can correct for both in-plane and through plane motion but are hampered by the need for additional scans and long reconstruction times [18]. Recently, DL has been used to speed up this process and, in some cases, dispense with the need for additional reference scans [19,20]. However, most methods focus on aligning ventricular blood pool segmentations rather than correcting the underlying cine images [21,22]. Furthermore, the magnitude of misalignment in free-breathing images is much greater than those resulting from breath-hold inconsistencies. Thus, we developed a method that can correct slice misalignment due to free-breathing without the need for a reference scan or pre-segmentation. We qualitatively demonstrated significant improvement in slice alignment after application of our DL model in prospective patients, with very short inference times (7 seconds for 32 volumes). The robustness of our method can be explained by several factors. Firstly, acquiring slices in the sagittal orientation limited respiratory movement mostly to the Head-Foot and Anterior-Posterior directions, allowing motion correction to be achieved through in-plane image deformations alone. This significantly simplified the realignment problem and enabled a relatively simple UNet architecture to be used. Secondly, we simulated respiratory motion using carefully crafted physiological non-rigid deformations that contained large amounts of misalignment. We believe this enabled robust correction even in patients with unusual breathing patterns. Finally, we calculated the losses on the final corrected volumes rather than the deformation fields themselves. This enforced deformation fields that optimised image quality in the 3D data rather than enforcing accurate estimation of the fields themselves.

**Slice based super-resolution and final image quality**

After motion correction, super-resolution in the slice direction was required. Slice-based DL super-resolution has previously been developed for both cine and late enhancement imaging with robust results [23]. In addition, DL super-resolution techniques have also been developed to recover in-plane detail in 2D and 3D CMR [24]. These studies leveraged varying DL approaches including fully supervised convolutional networks, self-supervised methods and generative adversarial networks (GANs) [25,26]. More recently, end-to-end DL pipelines for integrated motion correction and super-resolution have been developed that show superior performance to separate models [21].

We utilised a UNet architecture to perform super-resolution, but unlike previous studies we did not interpolate the low-resolution data before it was inputted into model. The theoretical benefit of this approach is that it allows the model to 'learn' the optimum interpolation and resolution recovery. Another modification was to train the super-resolution model in an end-to-end manner with a respiratory-correction model. This allowed the super-resolution model to also learn how to correct for any residual slice misalignment after motion correction, which improved final image quality. However, the respiratory-correction model trained in an end-to-end manner was inferior to the model trained independently. This was probably because of difficulties in balancing the weighting between the final loss and the intermediate loss (calculated after respiratory-correction). Thus, in the prospective study we combined the independently trained respiratory-correction model with the end-to-end trained super-resolution model. We demonstrated that super-resolution significantly improved subjective image quality with experienced clinicians rating them diagnostically accurate. This is in comparison to the low-resolution data that was rated diagnostically poor. Super-resolution also improved edge sharpness to the level of conventional 3D WH imaging. However, it should be noted that the conventional 3D WH and 2D cine data had significantly better image scores (good and excellent respectively). Our findings suggest that the final 3D-cine images could be used for clinical assessment, but the lower image quality compared to conventional methods might limit identification of more subtle abnormalities.

**Segmentation and clinical utility**

After creation of the isotropic 3D-cine data, we performed automatic DL segmentation of the atria, ventricles, aorta and pulmonary arteries on a volume-by-volume basis. Unlike the other DL models in our pipeline, we used a UNet3+ for segmentation rather than a more conventional UNet. This architecture includes full scale skip connection and allows better multi-scale feature fusion, providing a richer understanding of both fine details and global context. This aids in segmentation and we have previously used the UNet3+ architecture to successfully segment short-axis images in single ventricular patients with extremely heterogenous anatomy. Using this model, we showed minimal bias and reasonable agreement in both right and left ventricular volumes compared to reference-standard 2D cine imaging. We also demonstrated minimal bias and reasonable agreement for great vessel diameter measurement against reference-standard 3D WH imaging. These results were in-line with previous real-time [27] and 3D-cine methods [4,17,28]. Importantly, both volumetric and diameter data were produced automatically, which has the potential to vastly improve clinical workflows. An important further benefit of our 3D-cine technique is that it produces volume time curves for all cardiac chambers, as well as measures of shape that are of increasing interest in congenital heart disease [29].

**Limitations**

The main limitation of this study was the relatively small population in the validation cohort. However, this was primarily a proof-of-concept study to demonstrate the feasibility of our methodology. The next step will be a wider study (optimally multi-centre) that additionally assesses diagnostic capability, test-retest, and atrial volumes in a much larger cohort.

Another limitation of our methods is that the DL models are applied independently to each volumetric frame of the cine data. Although this means that temporal consistency is not enforced, this does not appear to be a significant problem in our small sample. Nevertheless, it could be problematic in more complex congenital heart disease and a 4D approach may be desirable if enough training data were available.

In this study, we also only compared conventional imaging and manual segmentation against our fully automated approach. Thus, we didn't fully investigate whether any errors arise from deficiencies in the image enhancement or segmentation. Although this could be achieved by manually segmenting the 3D-cine data, this task is not trivial and could introduce new errors (particularly around definition of the valves). Furthermore, we believe that without automated processing the clinical impact of our 3D-cine approach is limited.

## **Conclusion**

This paper demonstrates the potential of creating 3D-cine data from concatenated 2D real-time cine images, resulting in higher contrast than other current 3D-cine methods. We used a series of UNet-like DL architectures to motion-correct, super-resolve and segment our 3D-cine data. All our DL models have been trained with open-source data making it widely accessible to other clinical centers. Our method has short acquisition and reconstruction times with fully segmented data being available within 2 minutes. This makes our technique particularly appealing within busy clinical workflows. We have shown that ventricular volumes and vessel diameter measurements from automatically segmented data have reasonable agreement with reference-standard, high-resolution 2D cine and 3D WH techniques. Thus, we believe that this technique may help speed up CMR in clinical practice.

# Abbreviations

4CH – four-chamber

Ao – thoracic aorta

bSSFP – balanced steady state free precession

CGM – classification guided module

CLAHE – Contrast Limited Adaptive Histogram Equalization

CNN – convolutional neural network

CMR – cardiovascular magnetic resonance

DL – Deep Learning

EDV – end-diastolic volume

EF – ejection fraction

ES – edge sharpness

ESV – the end-systolic volume

GMAE – gradient mean absolute error

GANs – generative adversarial networks

HighRes – high-resolution

HVSMR – Whole-Heart and Great Vessel Segmentation from 3D Cardiovascular MRI in Congenital Heart Disease

LA – left atrium

LPA – left pulmonary arteries

LowRes – low-resolution

LowRes$_{resp}$ – low-resolution respiratory-artefacted volume

LowRes$_{resp+band}$ – low-resolution volume with respiratory artefacts and contrast differences

LV – left ventricle

MAE – combined mean absolute error

MMWHS – Multi-Modality Whole-Heart Segmentation

MPR – multi-planar reformats

MUSIC – Multiphasic Steady-State Imaging with Contrast Enhancement

PA – pulmonary artery

PCHD – paediatric and congenital heart disease

RA – right atrium

RPA – right pulmonary arteries

RV – right ventricle

SAX – short axis

SVR – slice-to-volume registration

TE – echo time

TR – repetition time

VCG – vectorcardiogram

WH – whole-heart

## **Acknowledgements**

This work was supported by UK Research and Innovation (EP/S021612/1, MR/S032290/1), the British Heart Foundation (FS/4yPhD/F/22/34181, FS/ICRF/22/26046). The funders had no role in study design, data collection and analysis, decision to publish, or preparation of the manuscript.

|  | 3D-cine | Reference-standard BH 2D cine | Bias | Limits of agreement | p-value |
|---|---|---|---|---|---|
| LV EDV (ml) | 116±38 | 117±36 | -0.4 | -15.4 to 14.6 | 0.6953 |
| LV ESV (ml) | 43±16 | 42±14 | -1.3 | -14.0 to 11.4 | 0.7695 |
| LV EF (%) | 63±7 | 64±6 | 0.9 | -5.7 to 7.6 | 0.6953 |
| RV EDV (ml) | 132±49 | 136±43 | 4.4 | -14.5 to 23.2 | 0.2324 |
| RV ESV (ml) | 58±35 | 53±25 | -3.7 | -32.1 to 24.6 | 0.4922 |
| RV EF (%) | 58±8 | 61±8 | 2.7 | -13.0 to 18.5 | 0.4316 |

Table 1: Functional cardiac parameters as calculated from 3D-cine and reference-standard breath-hold (BH) 2D cine. EDV - end-diastolic volume; ESV - end-systolic volume; EF - ejection fraction; for LV - left ventricle and RV - right ventricle. Metrics are displayed as mean ± standard deviation. Bias is the mean of the paired difference (Reference-standard BH 2D cine – 3D-cine) presented with limits of agreement (1.96*standard deviation of the difference). p-value from Wilcoxon signed-rank test.

|  | 3D-cine | Reference-standard 3D WH | Bias | Limits of agreement | p-value |
|---|---|---|---|---|---|
| **Ao (mm)** | 25.2±2.6 | 25.6±4.0 | 0.45 | -3.5 to 4.4 | 0.5145 |
| **MPA (mm)** | 23.3±3.3 | 22.9±3.7 | -0.44 | -5.1 to 4.2 | 0.5899 |
| **LPA (mm)** | 17.0±3.1 | 17.0±3.6 | 0.02 | -2.3 to 2.4 | 0.9610 |
| **RPA (mm)** | 16.3±2.3 | 15.5±2.2 | -0.80 | -2.0 to 0.4 | 0.0037 |

Table 2: Vessel diameter measurements from static 3D whole-hearts and 3D-cines. Ao – Aorta; MPA – Main pulmonary artery; LPA – Left pulmonary artery; RPA – Right pulmonary artery. Metrics are displayed as mean ± standard deviation. Bias is the mean of the paired difference (3D Whole-Heart – 3D-cine) presented with limits of agreement (1.96*standard deviation of the difference). p-value from paired t-test.

|  | ES (4CH) | ES (SAX) | Contrast (4CH) | Contrast (SAX) |
|---|---|---|---|---|
| **Uncorrected concatenated real-time volumes** | 0.26±0.06 | 0.28±0.06 | 3.4±1.0 | 3.0±0.66 |
| **3D-cine** | 0.34±0.04 ^ | 0.34±0.04 ^ | 3.3±0.98 | 3.0±0.54 |
| **Reference-standard 3D WH** | 0.36±0.03 ^ | 0.37±0.11 | 3.9±1.1 | 3.8±1.25 * |
| **Reference-standard BH 2D cine** | 0.45±0.06 ^,*,† | 0.42±0.04 ^,* | 8.5±1.9 ^,*,† | 8.2±1.56 ^,*,† |

Table 3: Mean ± standard deviation for edge sharpness (ES) and blood pool-myocardial contrast measurements for four chamber (4CH) and short axis (SAX) views. for prospective data. ^,*,† indicates statistical significance from uncorrected volumes, 3D-cine and Reference-standard 3D Whole-Heart and Reference-standard BH 2D cine images respectively.

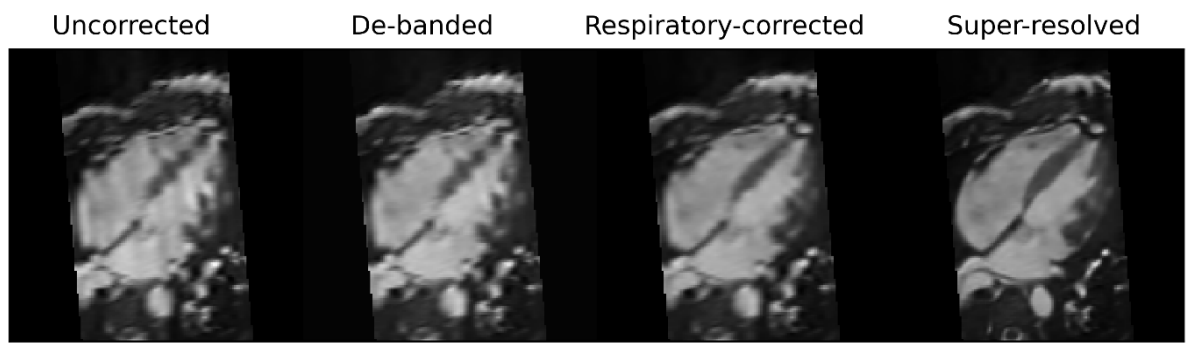

Figure 1: Shows improvement of uncorrected concatenated real-time data in four chamber (4CH) view after sequentially applying DL models

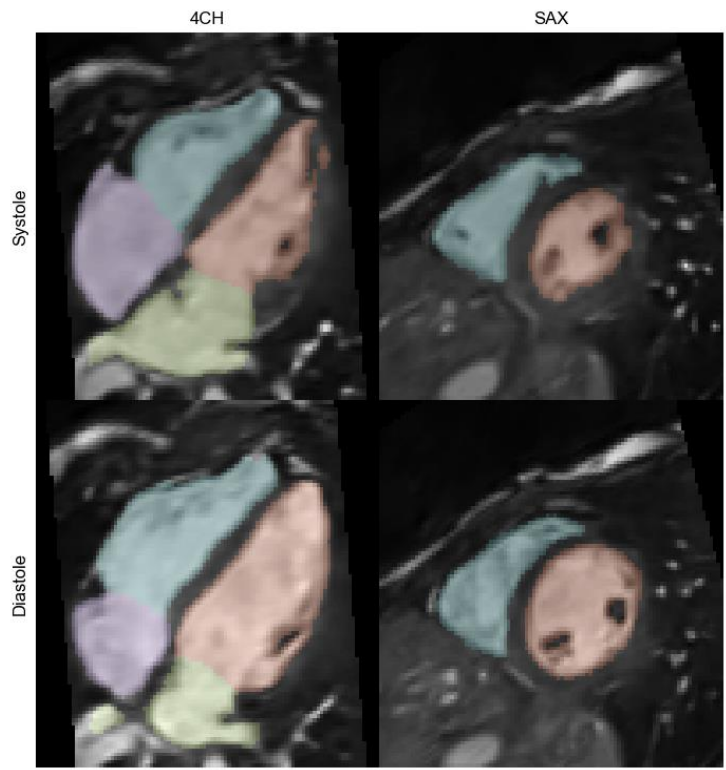

Figure 2: Shows example automatic DL segmentation from 3D-cine for the four chamber (4CH) and short axis (SAX) views in systole and diastole (Blue = Right ventricle, Red = Left ventricle, Purple = Right atrium, Green = Left atrium)

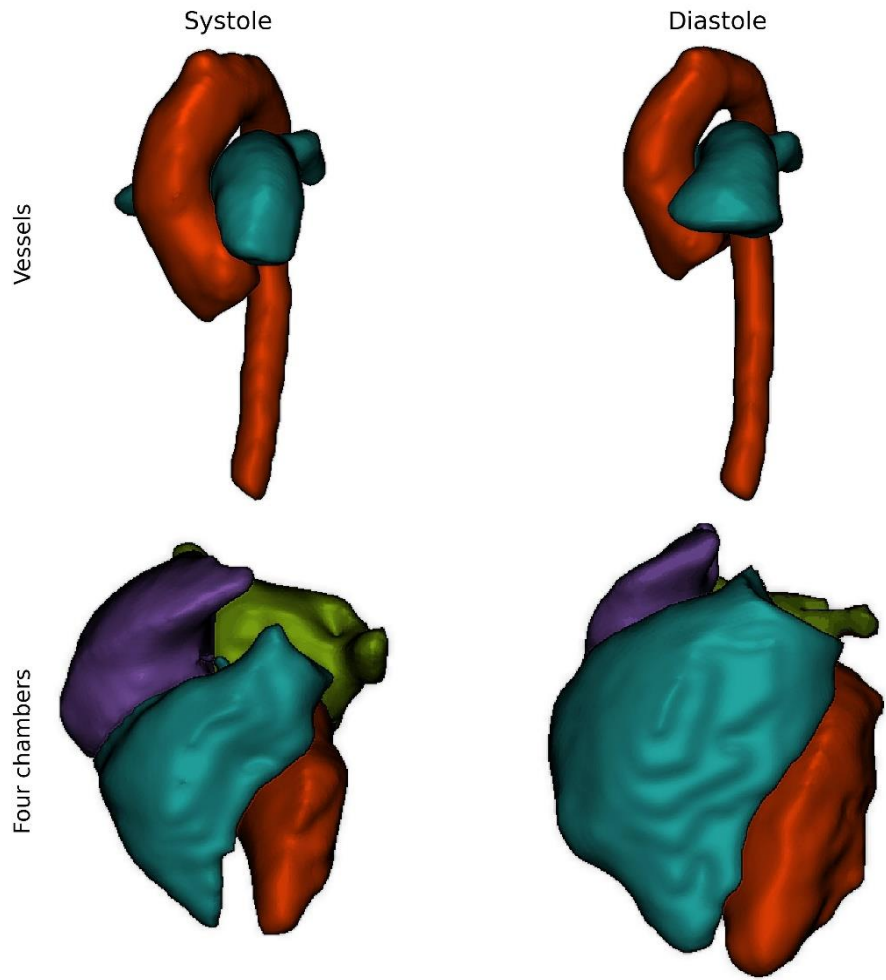

Figure 3: Shows a 3D render of the great vessels and four-chambers using automatic Deep Learning (DL) segmentation from 3D-cine at systole and end diastole. (Blue = Right ventricle, Red = Left Ventricle, Purple = Right atrium, Green = Left atrium, Vessels: Red = Aorta, Blue = Pulmonary arteries).

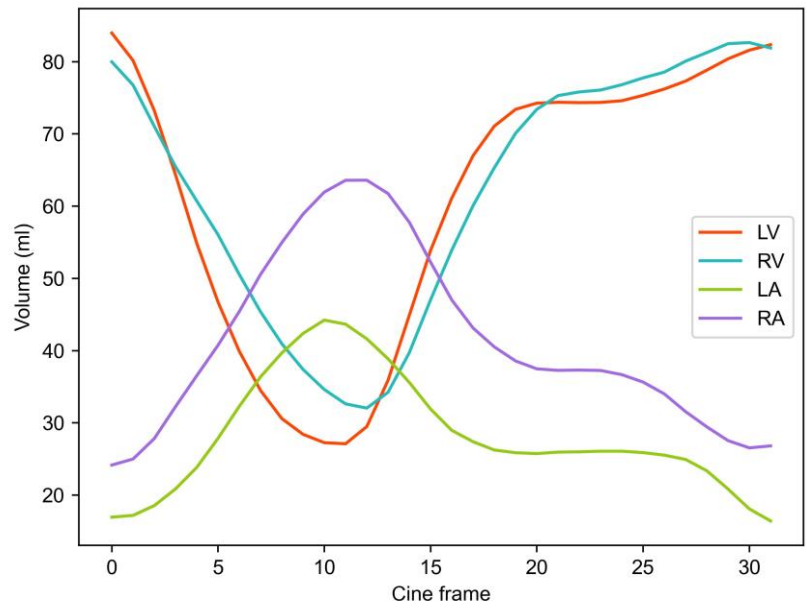

Figure 4: Shows example time-volume curves for the four-chambers, automatically extracted from the Deep Learning (DL) segmentation. LV – Left Ventricle, RV – Right Ventricle, LA – Left atrium, RA – Right atrium.

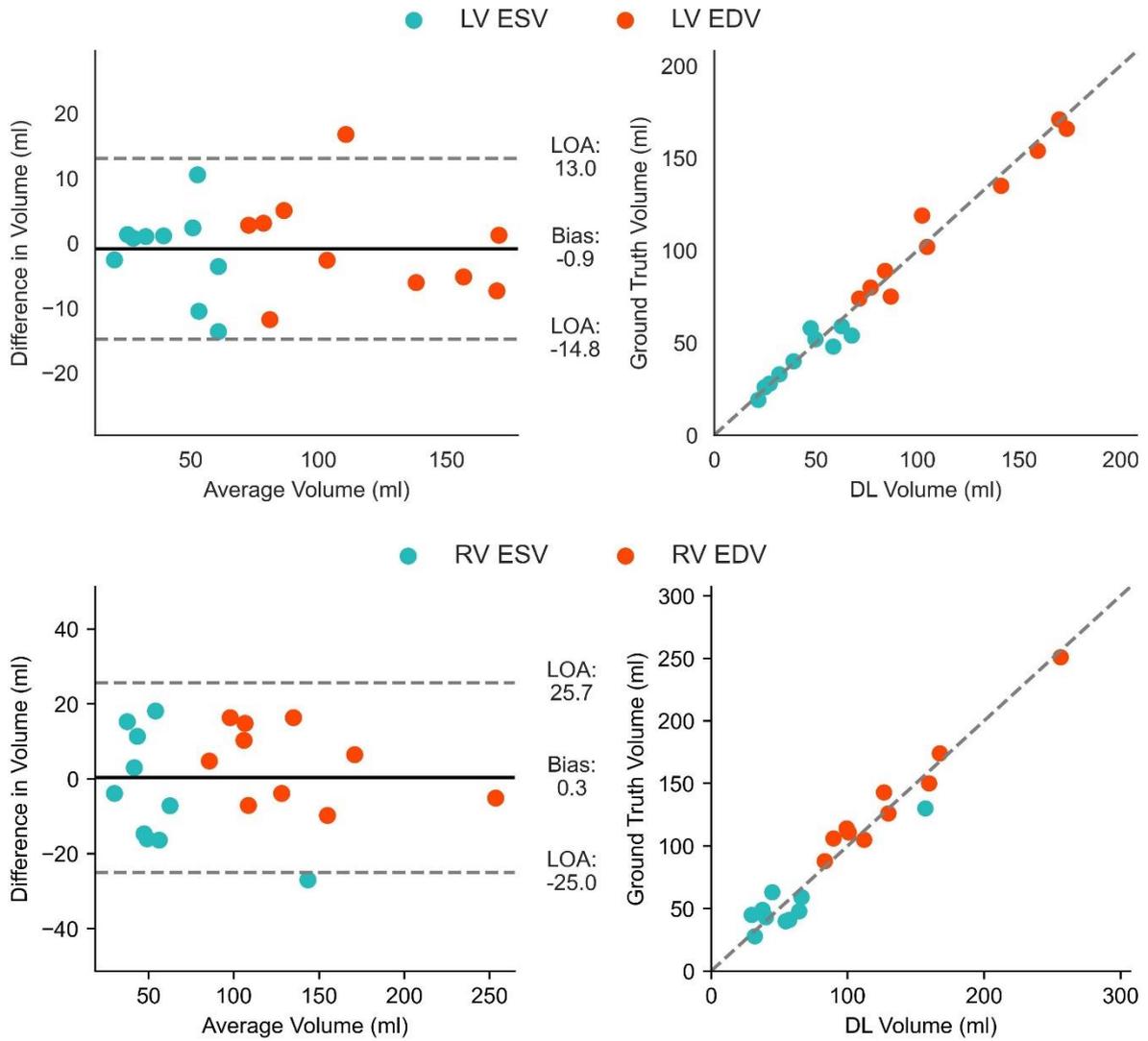

Figure 5: Shows Bland-Altman and scatter plots of right ventricular (RV) end systolic volume (ESV) and end diastolic volume (EDV) (top) and left ventricular (LV) ESV and EDV (bottom) for 3D-cine automatic segmentation against reference-standard BH 2D cines with manual segmentation.

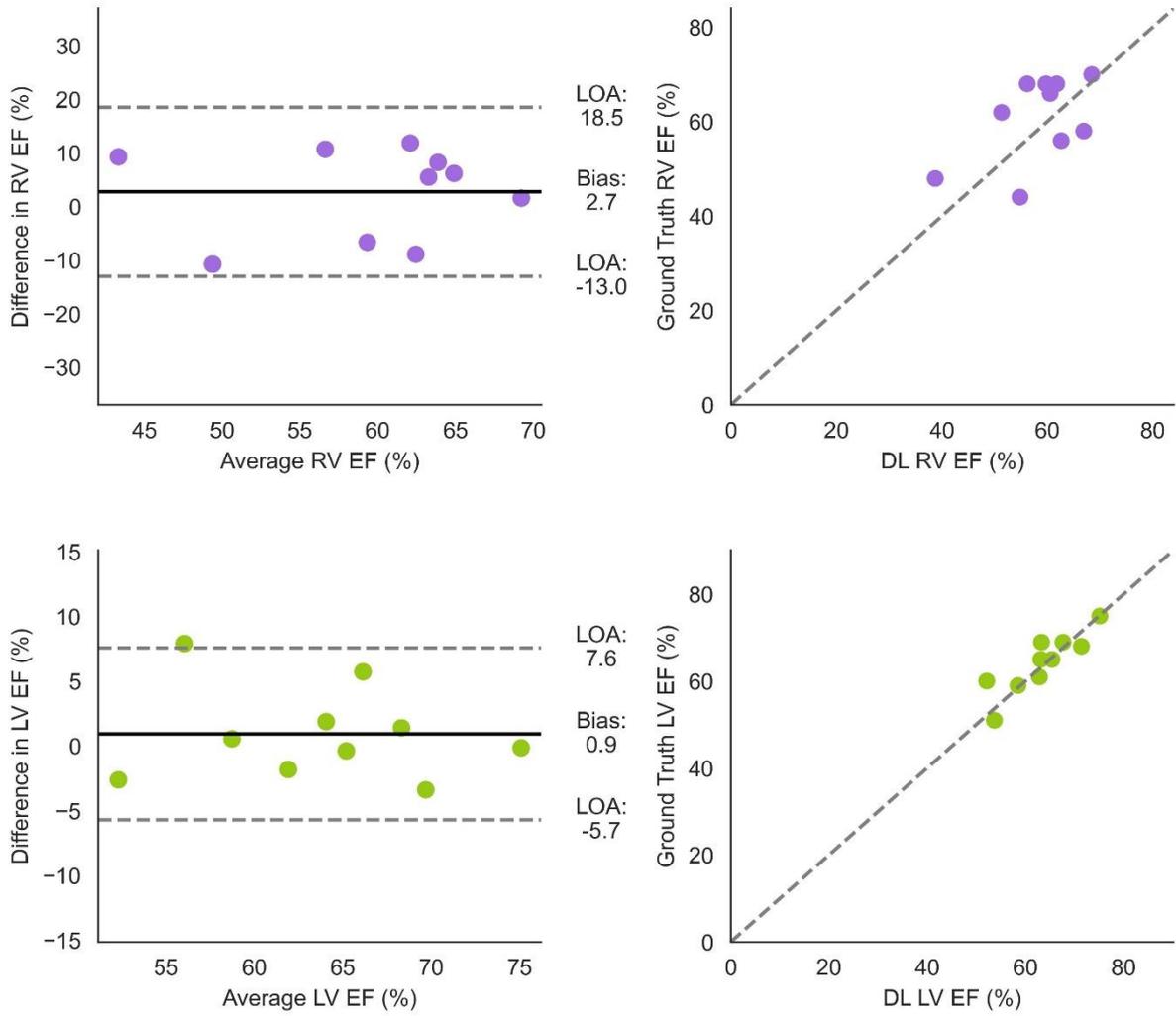

Figure 6: Shows Bland-Altman and scatter plots of right ventricular (RV) ejection fraction (EF) (top) and left ventricular (LV) EF (bottom) for 3D-cine automatic segmentation against reference-standard BH 2D cines with manual segmentation.

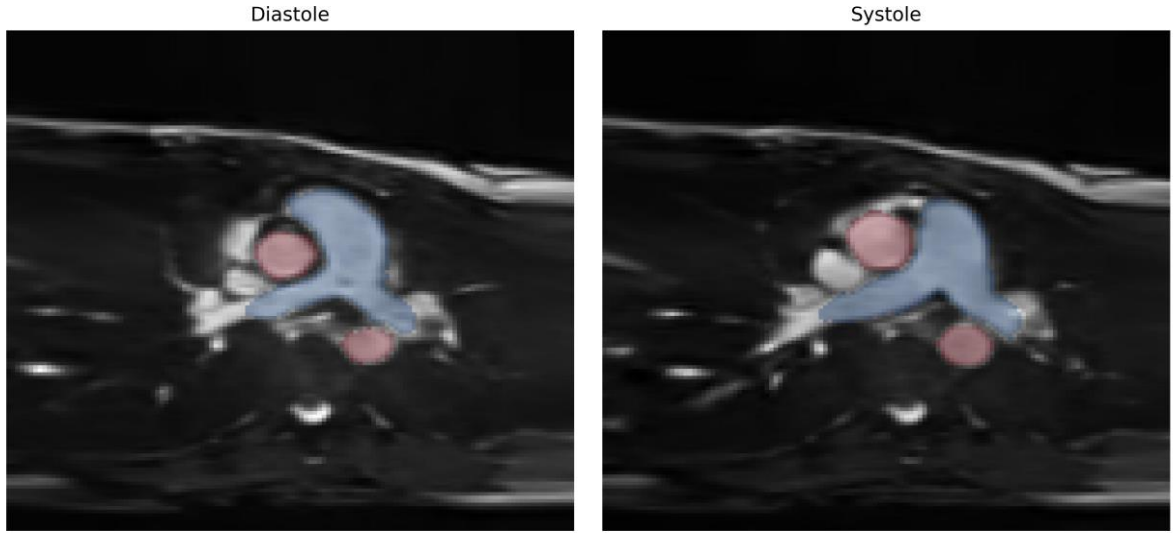

Figure 7: Shows automatic Deep Learning (DL) segmentation from 3D-cine of the aorta (Red) and pulmonary artery (Blue) from 3D-Cine data.

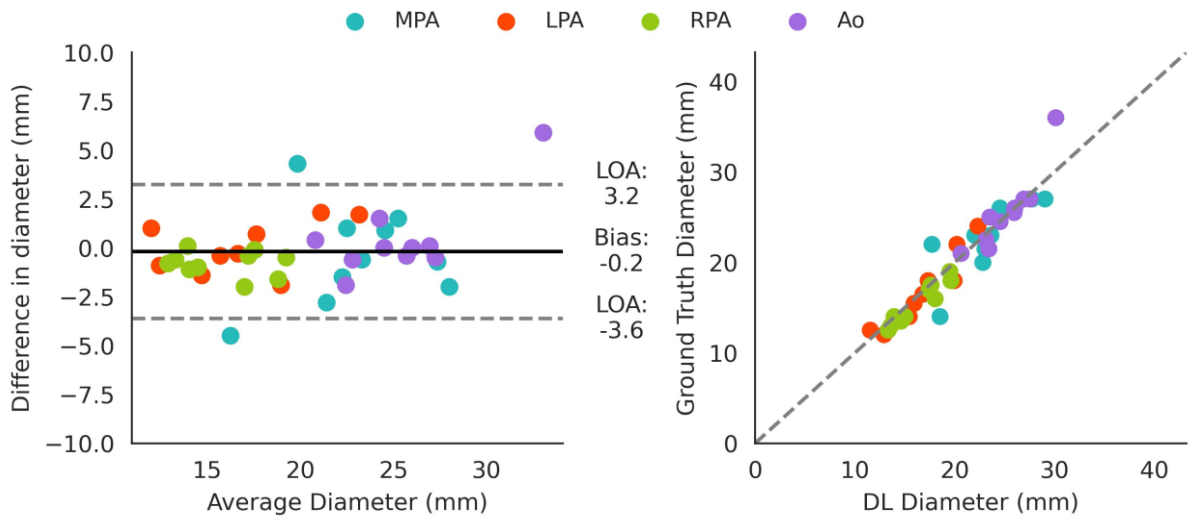

Figure 8: Shows a Bland-Altman and scatter plot aortic and main pulmonary artery sino-tubular junction (Ao and MPA) diameter, as well as the midpoint of the left and right pulmonary arteries (RPA and LPA) diameter measurements from 3D-cine automatic segmentation against reference-standard 3D whole-heart data with manual segmentation.

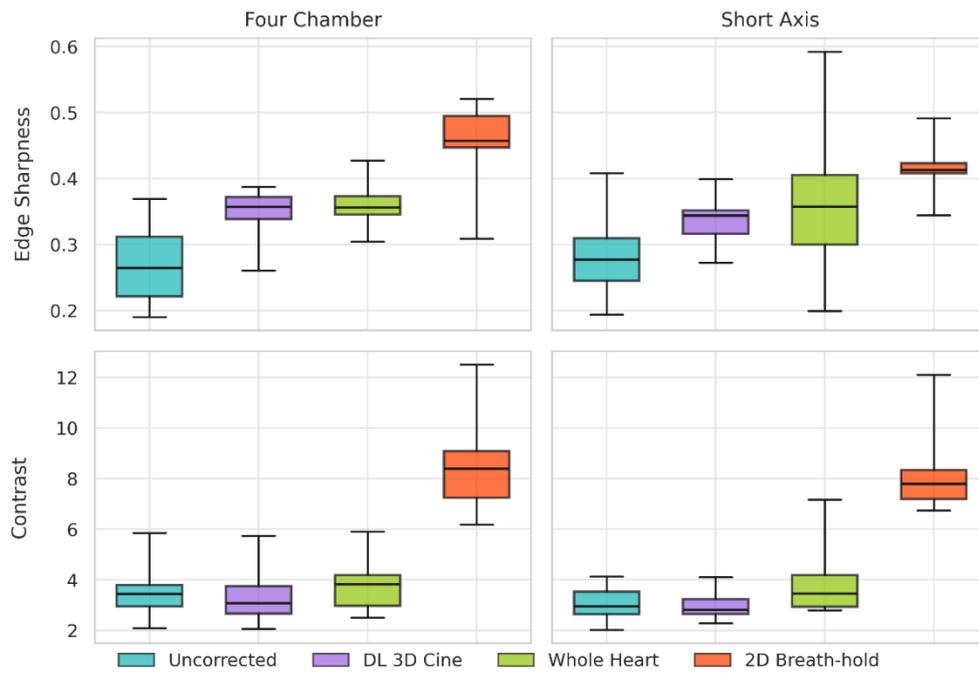

Figure 9: Shows Box plots of the edge sharpness and blood pool-myocardial contrast of the uncorrected *concatenated* real-time input data, the fully corrected 3D-cine data, the reference-standard 3D whole-heart and reference-standard BH 2D cine data for both four chamber and short axis views.

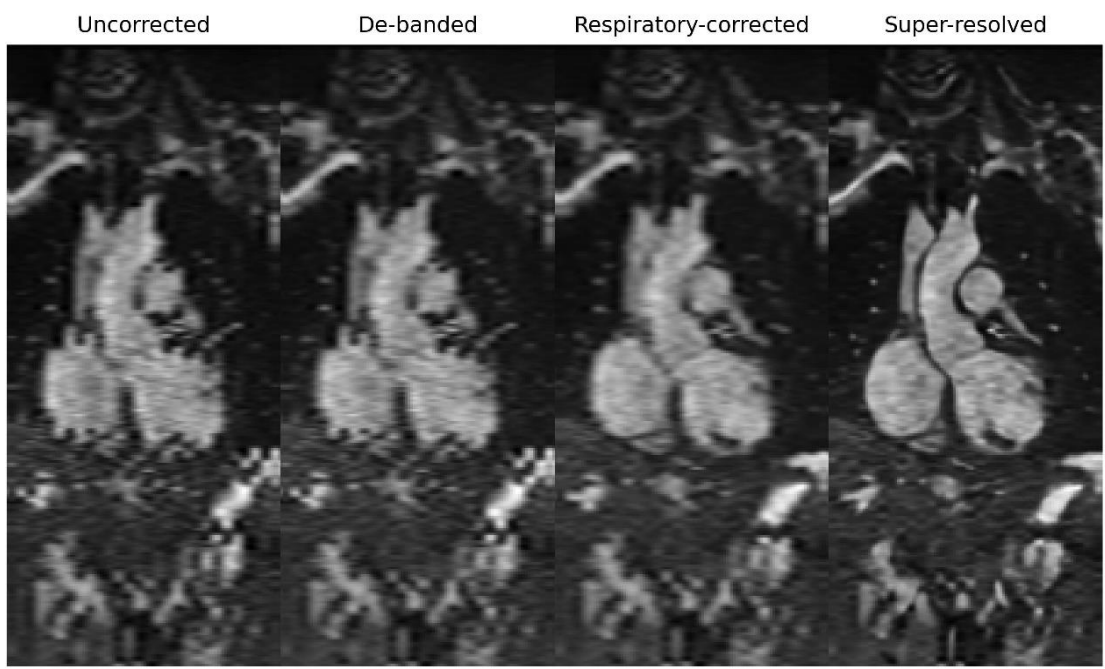

Figure 10: Shows improvement of uncorrected concatenated real-time data in four chamber (4CH) view after sequentially applying DL models

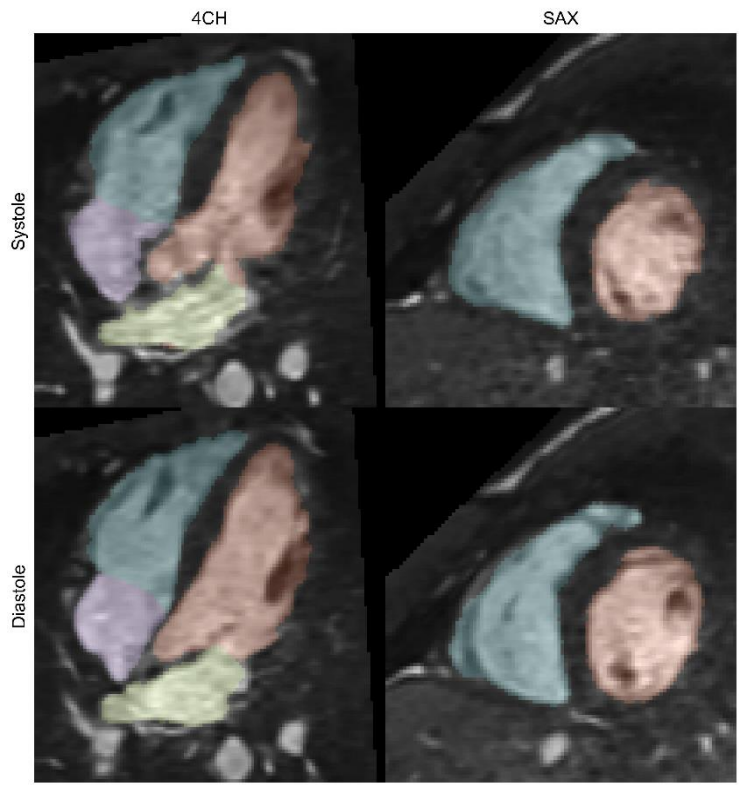

Figure 11: Shows example automatic DL segmentation from conventional real-time data for the four chamber (4CH) and short axis (SAX) views in systole and diastole (Blue = Right ventricle, Red = Left ventricle, Purple = Right atrium, Green = Left atrium)

# Supplement

## Supplemental Information 1: Volumentations for Augmentation of Segmentation Data

For training of the segmentation model, on-the-fly augmentations were performed using the volumentations library (version 1.0.4). Rotations were performed in all directions by up to ±10 degrees with a probability of 0.75 to be applied to any given dataset. Small random elastic transforms were also performed with a random magnitude ranging between 0-0.1, with 0.75 probability of application.

## Supplemental Information 2: Hyperband Optimisation for Image Correction Networks

The image correction networks (de-banding, respiratory-correction and super-resolution) used an image-based loss which combined mean absolute error (MAE) and gradient mean absolute error (GMAE). The ratio between MAE and GMAE losses was optimised using Hyperparameter optimization (keras Hyperband), where the GMAE loss was varied between 0.1-10 relative to the MAE loss. The loss used for optimization was a standard mean square error (MSE) loss and the optimization was run for 10 epochs. The hyperband optimized ratio of GMAE to MAE loss was 5.357:1.

## Supplemental Information 3: Regularization Loss for Respiratory-correction Network

Training of the respiratory correction network used the Hyperband optimised MAE and GMAE image-based loss function, with an additional regularization loss. This regularization loss was applied to the in-slice deformation fields to promote smooth, realistic corrective deformations. This loss is calculated by computing the 2D Sobel gradient for each slice on the x and y deformation fields. The squared L2 norm is calculated over the x and y gradient components, summed over all the slices and averaged. This regularization loss is then scaled by $5 \times 10^{-8}$ and combined with the image-based losses (MAE and GMAE).

## Supplemental Information 4: Loss Function for Segmentation Network

A surface area loss is used in combination with a Focal Tversky loss during the training of the segmentation model for the RA, LA, RV, LV, PA and Ao. The surface area loss compares the number of pixels on the surface of the masks between the predicted and ground truth segmentations. For each structure of interest, it converts the corresponding binary mask to a surface by first creating an element-wise negative of the input mask. The negative mask is then passed through a MaxPool3D (tensorflow function with kernel size = 3, strides = 1) which effectively erodes the structure of interest. By subtracting the negative mask from the eroded mask, we are left with a thin, hollow surface around the structure of interest. The pixels on this surface are then summed to give a scalar value, proportional to the surface area of the structure of interest. The surface area loss is then calculated as the square difference of the predicted surface area and the ground truth surface area, summed across all six segmented structures. The surface area loss is finally combined with the Focal Tversky loss, with a surface area scaling factor of $7.62 \times 10^{-6}$.

|  | LV | RV | LA | RA | Ao | PA |
|---|---|---|---|---|---|---|
| **Train** | 0.88±0.11 | 0.83±0.20 | 0.81±0.09 | 0.82±0.22 | 0.83±0.16 | 0.83±0.13 |
| **Validation** | 0.83±0.12 | 0.73±0.29 | 0.68±0.24 | 0.72±0.29 | 0.78±0.10 | 0.73±0.18 |

Supplementary Table 1: Dice scores for each segmented structure for training and validation datasets. LV – Left Ventricle, RV – Right Ventricle, LA – Left atrium, RA – Right atrium, Ao - Aorta , PA – Pulmonary Artery

|  | Uncorrected volumes | De-banded volumes | Respiratory-corrected volumes | Super-resolved volumes |
|---|---|---|---|---|
| **Diastolic axial slices** | 3.3±0.9 | 3.0±0.7 | 2.2±0.4 ^ | 1.0±0.0 ^,*,† |
| **4CH/SAX MPRs** | 3.3±0.9 | 2.8±0.8 | 2.1±0.6 ^,* | 1.1±0.6 ^,*,† |

Supplementary Table 2: Mean values and standard deviations for image quality ranking of prospective data represented as a stack of diastolic axial slices and four chamber / short axis (4CH/SAX) multi-planar reformats (MPRs) as different stages of sequential Deep Learning (DL) correction. ^,*,† indicates statistical significance in image quality from uncorrected, de-banded and respiratory-corrected volumes respectively.

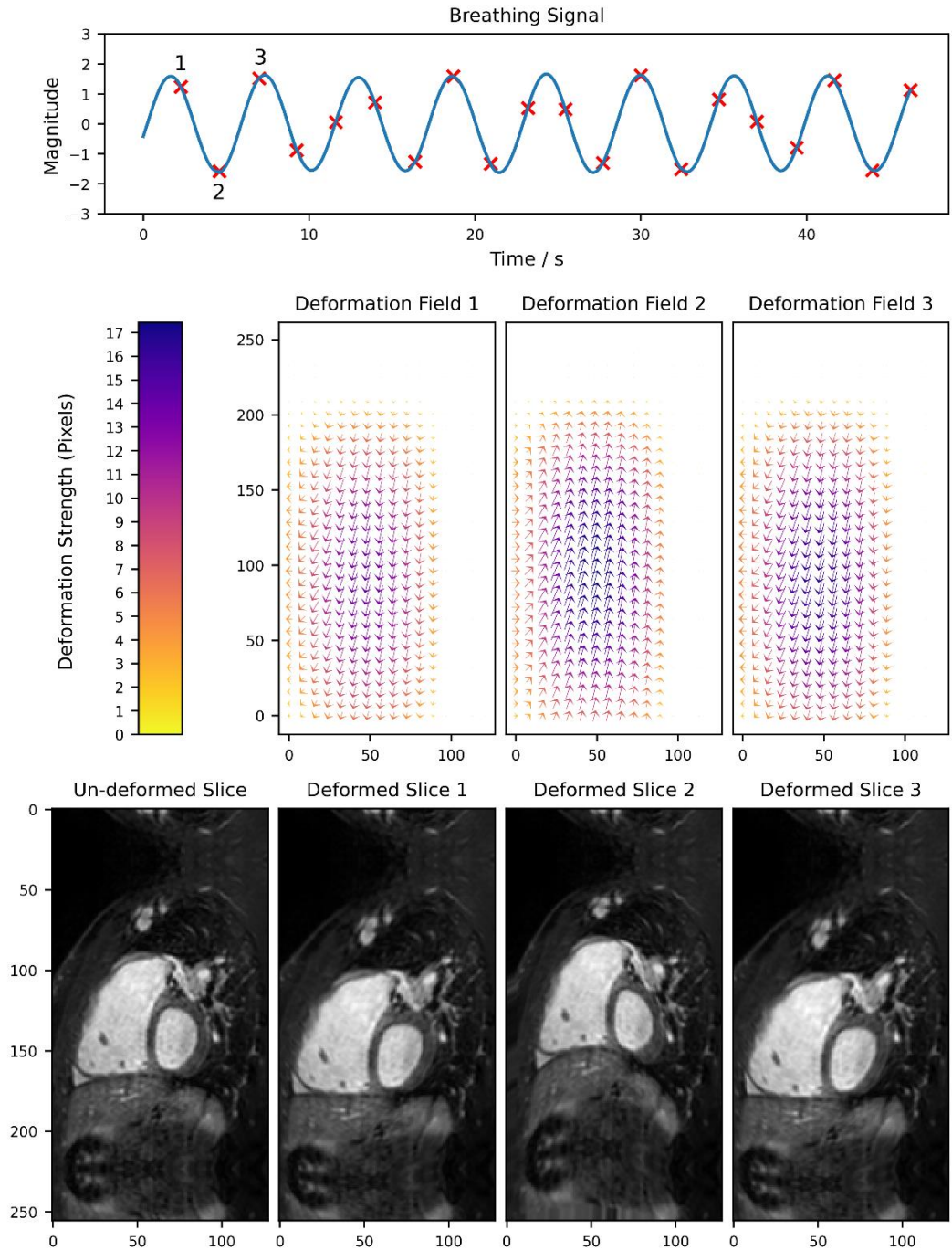

Supplementary Figure 1: Shows method of applying synthetic respiratory artefacts to data. Top = Synthetic respiratory signal and the sampling points corresponding to each slice, followed by an undeformed example slice. Middle = Three example deformation fields created from sampling the first three points. Bottom = The example slice after being deformed by the three middle deformation fields.

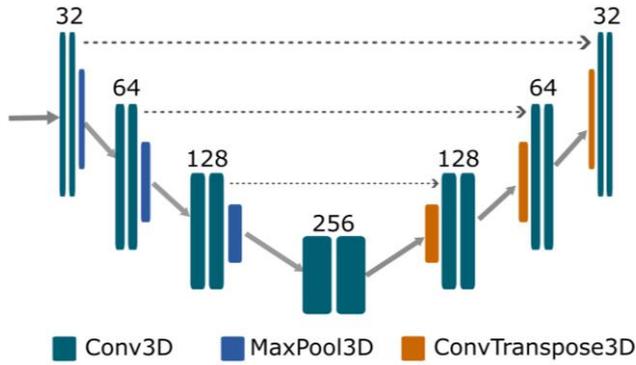

Supplementary Figure 2: Shows standard UNet3D architecture used for de-banding model with the number of filters shown at each layer depth

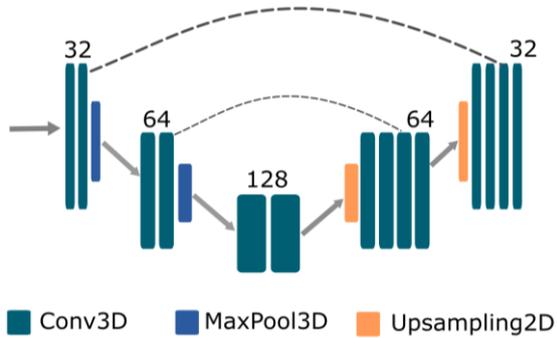

Supplementary Figure 3: Shows modified UNet3D architecture used for respiratory-correction model with the number of filters shown at each layer depth

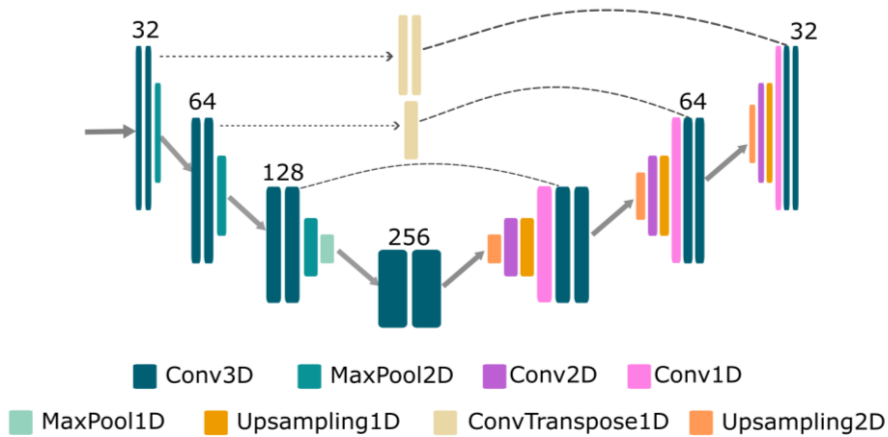

Supplementary Figure 4: Shows modified UNet3D architecture used for super-resolution model with the number of filters shown at each layer depth

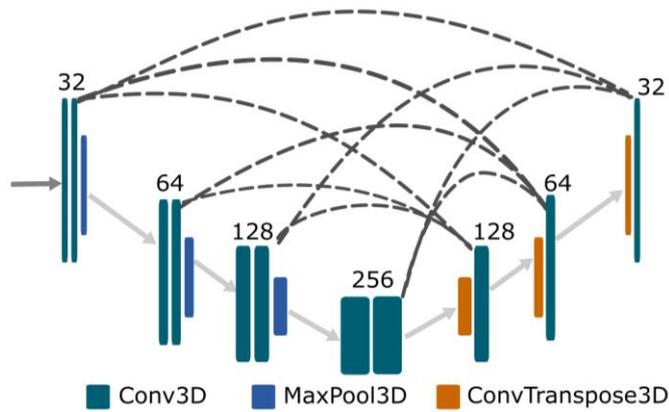

Supplementary Figure 5: Shows modified UNet3+ architecture used for segmentation with the number of filters shown at each layer depth

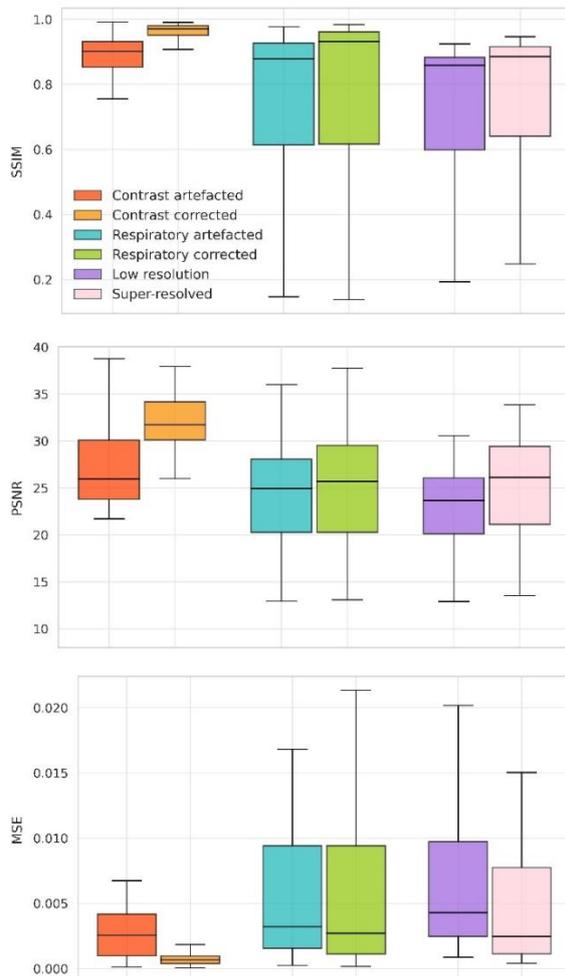

Supplementary Figure 6: Shows box plot comparing the value of image quality metrics (SSIM = top, PSNR = Middle, MSE = bottom) on validation data, before and after application of DL model for each stage of correction

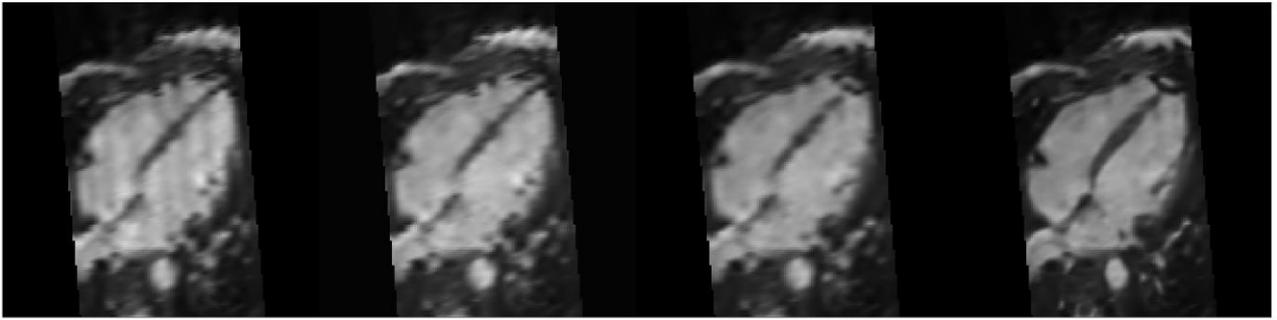

Supplementary Movie 1: Shows improvement of uncorrected real-time data in four chamber (4CH) view after sequentially applying Deep Learning (DL) models (left = Uncorrected, middle-left = de-banded, middle-right = respiratory corrected, right = Super resolved)

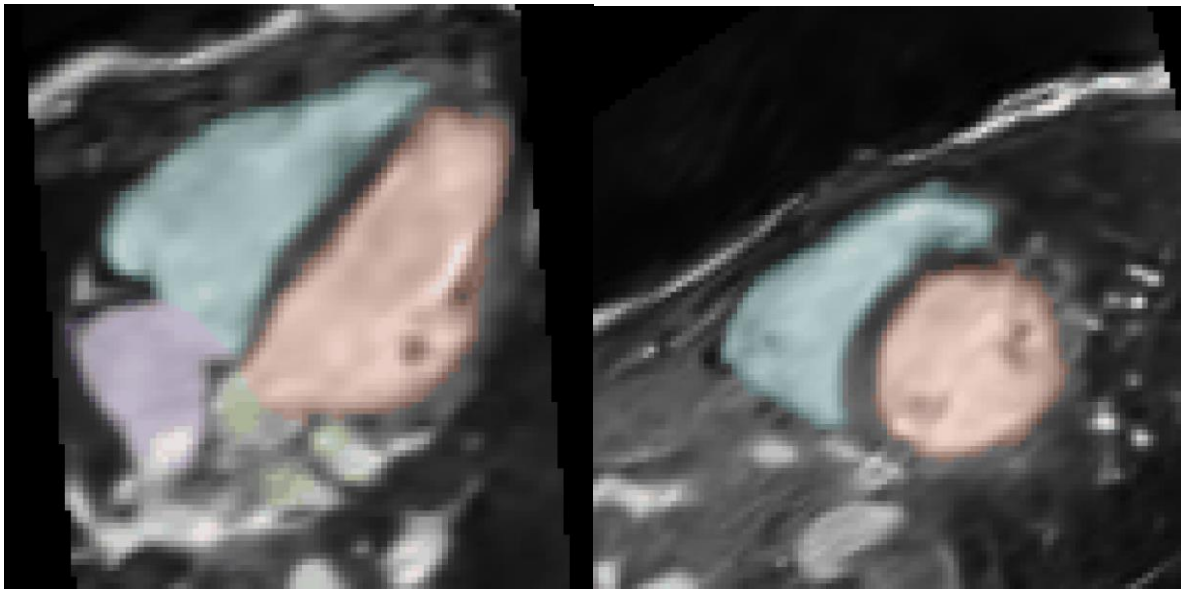

Supplementary Movie 2: Shows example 4 chamber (4CH) segmentations of our Deep Learning (DL) corrected data for the four chamber (4CH) (left) and short-axis (SAX) (right) over time (Blue – Right Ventricle, Red – Left Ventricle, Purple = Left Atrium, Green = Right Atrium)

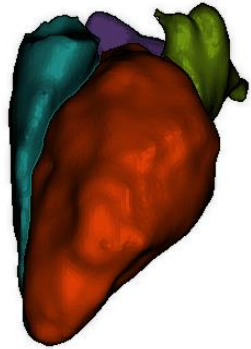

Supplementary Movie 3: Shows a 3D render of the four-chambers (4CH) over time (Blue – Right Ventricle, Red – Left Ventricle, Purple = Left Atrium, Green = Right Atrium)

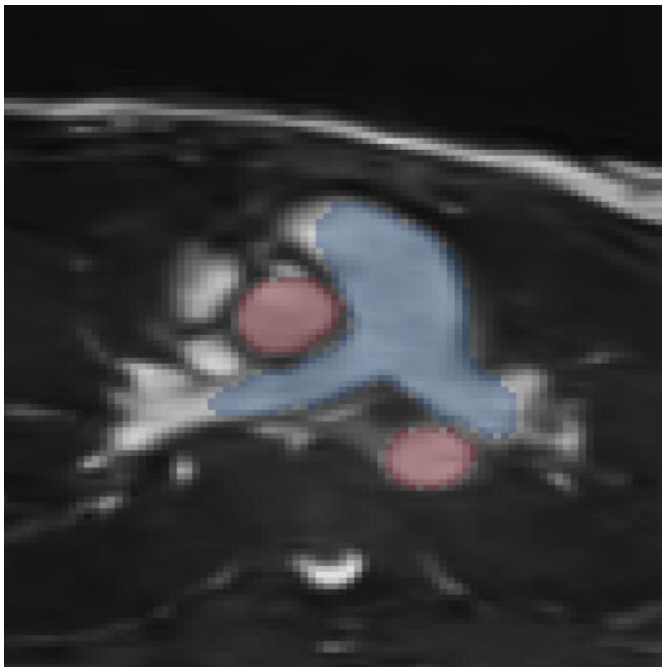

Supplementary Movie 4: Shows automatic Deep Learning (DL) segmentation of the aorta (Red) and pulmonary artery (blue) from 3D-Cine data over time

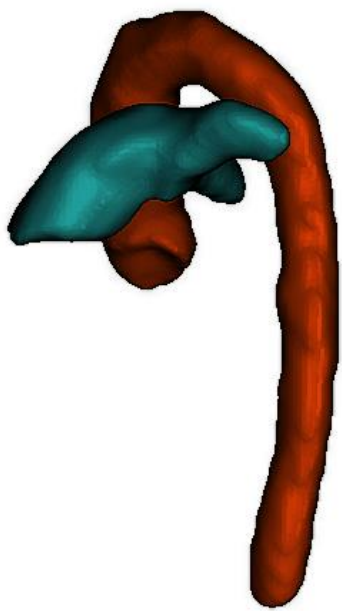

Supplementary Movie 5: Shows a 3D render of great vessels over time (Red = Aorta, blue = Pulmonary artery)

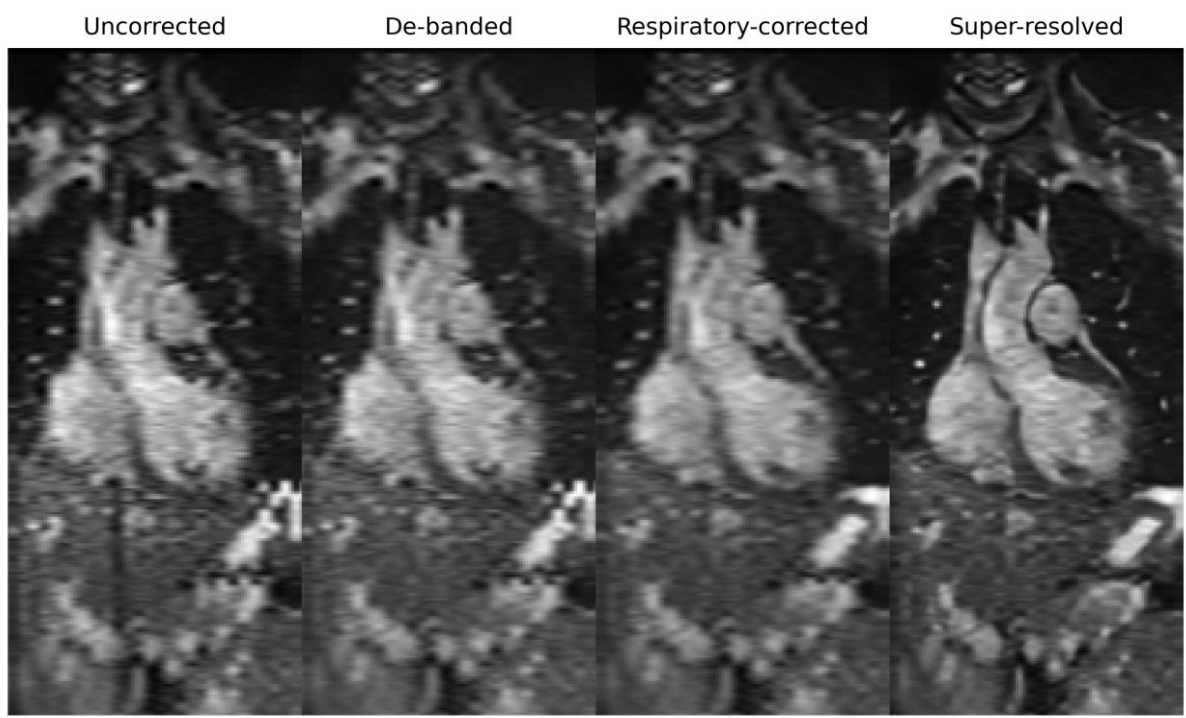

Supplementary Movie 6: Shows improvement of uncorrected conventional real-time data in coronal view after sequentially applying Deep Learning (DL) models.

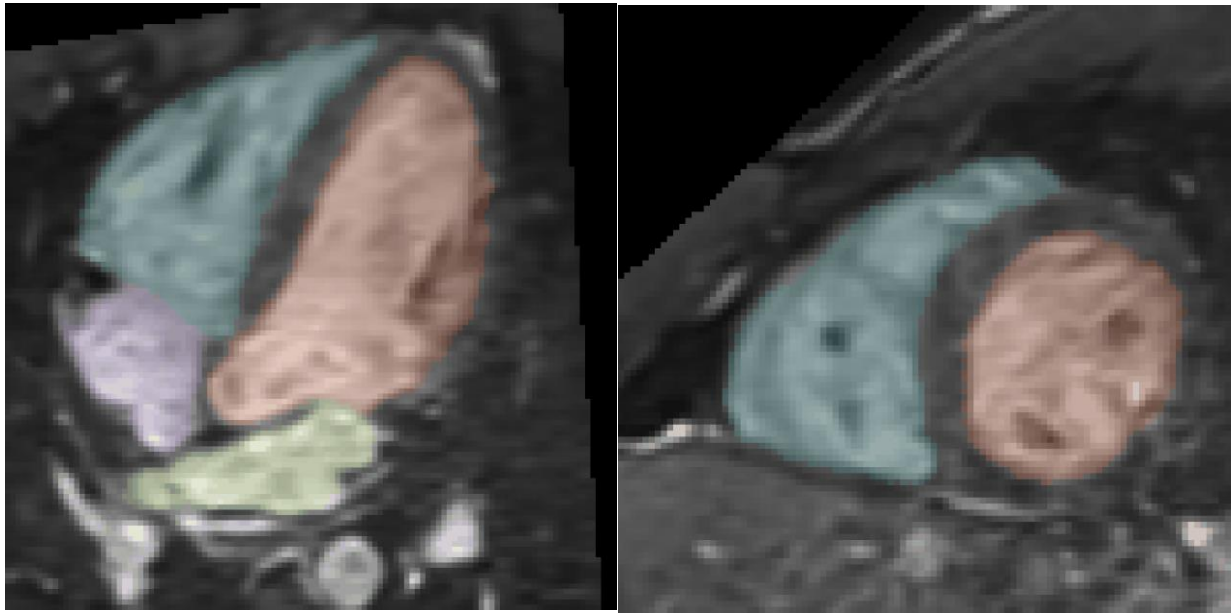

Supplementary Movie 7: Shows example segmentations of conventional real-time data for the four chamber (4CH) (left) and short-axis (SAX) (right) over time (Blue – Right Ventricle, Red – Left Ventricle, Purple = Left Atrium, Green = Right Atrium)